\documentstyle[12pt]{article}

\newcommand{\bce}{\begin{center}}
\newcommand{\ece}{\end{center}}
\newcommand{\beq}{\begin{equation}}
\newcommand{\eeq}{\end{equation}}
\newcommand{\bea}{\vspace{0.25cm}\begin{eqnarray}}
\newcommand{\eea}{\end{eqnarray}}

\newcommand{\ba}{\begin{array}}
\newcommand{\ea}{\end{array}}

\newcommand{\doublespace}{
    \renewcommand{\baselinestretch}{1.6}\large\normalsize}

\def\lsim{\mathrel{\rlap{\lower4pt\hbox{\hskip1pt$\sim$}}
    \raise1pt\hbox{$<$}}}         
\def\gsim{\mathrel{\rlap{\lower4pt\hbox{\hskip1pt$\sim$}}
    \raise1pt\hbox{$>$}}}         

\def\Pom{{\bf I\!P}}

\def\beq{\begin{equation}}
\def\endeq{\end{equation}}
\def\arr{\begin{eqnarray}}
\def\endarr{\end{eqnarray}}
\makeindex

  \advance\oddsidemargin  by -1in
  \advance\evensidemargin by -1in
  \advance\topmargin      by -0.5in
\begin{document}

\phantom{.}{\large \bf \hspace{8.4cm} KFA-IKP(Th)-1994-36 \\
\phantom{.}\hspace{10.1cm} DFTT 41/94 \\
\phantom{.}\hspace{9.9cm}3 October    1994\vspace{.4cm}\\
}

\begin{center}
{\Large \bf Direct calculation of the triple-pomeron coupling \\
for diffractive DIS
 and real photoproduction
\vspace{1.0cm}\\}
{\large \bf M.Genovese$^{a}$, N.N.~Nikolaev$^{b,c}$,
B.G.~Zakharov$^{b,c}$ \vspace{1.0cm}\\}
{\it
$^{a}$ Dipartimento di Fisica Teorica, Universit\`a di Torino,\\
and INFN, Sezione di Torino, Via P.Giuria 1, I-10125 Torino, Italy
\medskip\\
$^{b}$IKP(Theorie), KFA J{\"u}lich, 5170 J{\"u}lich, Germany
\medskip\\
$^{c}$L. D. Landau Institute for Theoretical Physics, GSP-1,
117940, \\
ul. Kosygina 2, Moscow V-334, Russia.\vspace{1.0cm}\\ }
{\Large \bf Abstract }\\
\end{center}
We present a unified direct calculation of the triple-pomeron
coupling $A_{3\Pom}(Q^{2})$ for
diffractive real photo\-pro\-duction ($Q^{2}=
0$) and deep inelastic scattering at large $Q^{2}$
in the framework of the dipole approach
to the generalized BFKL pomeron.
The small
phenomenological value of $A_{3\Pom}(0)\approx 0.16$\,GeV$^{2}$,
which was
a mystery, is related to the small correlation
radius $R_{c}\approx 0.3$\,fm for the perturbative gluons. We
confirm the early expectations of weak $Q^{2}$ dependence of
the dimensionfull coupling $A_{3\Pom}(Q^{2})$ and predict
that it rises by the
factor $\sim 1.6$  from real photoproduction to deep inelastic
scattering.
\bigskip\\

\begin{center}
E-mail: kph154@zam001.zam.kfa-juelich.de
\end{center}

 \doublespace
\pagebreak


\section{Introduction}

Salient feature of diffraction dissociation
$a+p\rightarrow X+p'$ of ($a=h$) hadrons and ($a=\gamma$)
real photons ($Q^{2}=0$) is the so called triple-pomeron regime
\beq
{M^{2} \over \sigma_{tot}(ap) } \cdot
\left.{d\sigma_{D}(a\rightarrow X)\over dt dM^{2}}\right|_{t=0}
\approx A_{3\Pom}\, .
\label{eq:1.1}
\endeq
Here $t$ is the $(p,p')$ momentum transfer squared, and
the mass $M$ of the excited state satisfies $m_{p}^{2} \ll
M^{2} \ll W^{2}$, where $W$ is the total {\sl c.m.s.} energy.
Eq.~(\ref{eq:1.1}) with the approximately energy- and
projectile-independent triple-pomeron coupling
$A_{3\Pom}$, holds at moderately
high energies, such that the photoabsorption and hadronic cross
sections are approximately constant ([1], for a review see [2]).
Notice that $A_{3\Pom}$ is a dimensional coupling:
$[A_{3\Pom}]=[{\rm GeV}]^{-2}$. The
Fermilab data on the diffractive real photoproduction give
\footnote{Apart from $\approx 5\%$ statistical error and $16\%$
normalization uncertainty [3], this number contains $\lsim 10\%$
uncertainty from our extrapolation from $|t|=0.05$\,GeV$^{2}$
to $t=0$ using the slope of the $t$-dependence as measured in [3].}
$A_{3\Pom}(Q^{2}=0) \approx 0.16$\,GeV$^{-2}$ [3]. In the Regge
theory language, the inclusive cross section of diffraction
dissociation measures the projectile-pomeron cross section [1]
\beq
 \sigma_{tot}(a\Pom;M^{2}\gg m_{p}^{2})={16\pi M^{2}
 \over \sigma_{tot}(pp) } \cdot
\left.{d\sigma_{D}(a\rightarrow X)\over dt dM^{2}}\right|_{t=0}
\approx 16\pi A_{3\Pom}{\sigma_{tot}(ap)\over \sigma_{tot}(pp)}\, .
\label{eq:1.2}
\endeq
Why $A_{3\Pom}$ is numerically small, and why the hadron-pomeron
cross section $\sigma_{tot}(a\Pom)$ is more than one order in
magnitude smaller than the hadron-nucleon cross section,
is one of outstanding mysteries of the pomeron.

The triple-pomeron regime will soon be explored
in details in an entirely new
domain of diffractive deep inelastic scattering (DIS) at HERA.
Here the underlying process is a diffraction
dissociation  of the virtual photon,
\beq
\gamma^{*}+p\rightarrow X+p'\, ,
\label{eq:1.3}
\endeq
at $x=Q^{2}/(Q^{2}+W^{2}) \ll 1$, where $Q^{2}$ is the virtuality
of the photon. The variable $x_{\Pom}=(M^{2}+Q^{2})/(W^{2}+Q^{2})
\ll 1$ can be interpreted as the fraction of proton's momentum taken
away by the pomeron, whereas $\beta = Q^{2}/(Q^{2}+M^{2})$ is the
Bjorken variable for DIS on the pomeron. Notice that
\beq
x_{\Pom}\beta = x \, .
\label{eq:1.4}
\endeq
The final-state proton $p'$ carries the fraction $(1-x_{\Pom})$ of
the beam proton's momentum and is separated from the hadronic debris
$X$ of the photon by a large
 (pseudo)rapidity gap $\Delta \eta \approx
\log{1\over x_{\Pom}}\gg 1$. In real photoproduction and
hadronic interactions, the pomeron exchange
was shown to dominate at $x_{\Pom} \lsim x_{\Pom}^{c} =$(0.05-0.1)
and/or $\Delta\eta \gsim \Delta \eta_{c}=$(2.5-3) [1,2]. In hadronic
interactions and/or real photoproduction, the triple-pomeron regime
corresponds to high c.m.s. energy of the $a\Pom$ interaction,
$M^{2} \gg m_{p}^{2}$, in DIS it requires $\beta \ll 1$ and/or
$M^{2} \gg Q^{2}$. For the evaluation of
$A_{3\Pom}(Q^{2})$ for DIS, one must consider
the moderately large rapidity gap such that
$x_{\Pom}\sim x_{\Pom}^{0}\lsim x_{\Pom}^{c}$, and
the moderately small
$x$, such that the proton structure function $F_{2}^{p}(x,Q^{2})$
is still approximately flat {\sl vs.} ${1\over x}$.
For reference point,
we will consider diffractive DIS at $x_{\Pom}=x_{\Pom}^{0}=0.03$.
Then, one can define
(for more precise definition of the related kinematical domain
see below)
\beq
{M^{2}+Q^{2} \over \sigma_{tot}(\gamma^{*}p)} \cdot
\left. { d\sigma_{D}(\gamma^{*}\rightarrow X)
 \over dt\,d M^{2} }\right|_{t=0} \approx
A_{3\Pom}(Q^{2}).
\label{eq:1.5}
\endeq
In diffractive DIS, the triple-pomeron coupling $A_{3\Pom}(Q^{2})$
controls [4] the normalization of the gluon and sea content of the
pomeron structure function.

The subject of this paper is a direct calculation of $A_{3\Pom}
(Q^{2})$ starting from the microscopic QCD description [4-6] of
diffraction dissociation in the framework of the (generalized)
dipole BFKL pomeron [5,6,8,9]. In section 2 we briefly review how
reaction (\ref{eq:1.3}) is described in terms of the diffraction
excitation of multiparton Fock states of the photon, which interact
with the target proton by the dipole BFKL pomeron exchange.
In section 3 we show that in DIS, the virtual photon
acts as an effective two-gluon (color octet-octet dipole)
state with the size of the order
of the correlation radius $R_{c}$ for the perturbative gluons.
We demonstrate how
the small $R_{c}\sim 0.3$\,fm, as suggested by lattice QCD
studies (for the recent review see [10]),
gives a natural small scale for $A_{3\Pom}(Q^{2})$. The
case of real photoproduction is studied in section 4. Here
the underlying mechanism of diffraction dissociation into large
masses is an excitation of $qg$ ($\bar{q}g$)
`clusters' ("constituent" quarks) of
size $R_{c}$, and we find $A_{3\Pom}(0)$
which agrees well with the experimental determination. This is
the first direct calculation of $A_{3\Pom}$ and the first
instance, when DIS and real photoproduction processes are shown
to share the dimensionfull coupling,
$[A_{3\Pom}]=[{\rm GeV}]^{-2}$,
which does not scale with $1/Q^{2}$, confirming earlier
conjectures [4-7]. Furthermore, in the scenario
[11,12] for the dipole cross section, we predict a slight, by the
factor $\sim 1.6$,  rise of
$A_{3\Pom}(Q^{2})$ from real photoproduction to DIS.
In section 5 we summarize our main results.


\section{Dipole pomeron description of the
diffraction excitation of photons}


We rely upon the microscopic QCD
approach to diffractive DIS developed in [4-7].
Diffraction excitation of the lowest $q\bar{q}$ Fock state of
the photon has the cross section (hereafter we focus on the
dominant diffraction dissociation of transverse photons)
\arr
\left.{d\sigma_{D}(\gamma^{*}\rightarrow X)\over dt}
\right|_{t=0}=
\int dM^{2}\,
\left.{d\sigma_{D}(\gamma^{*}\rightarrow X)\over dtdM^{2}}
\right|_{t=0}~~~~~~~~~~~~~~~~~~~~~~~\nonumber \\
= {1 \over 16\pi}
\int_{0}^{1} dz\int d^{2}\vec{r}\,\,
\vert\Psi_{\gamma^{*}}(Q^{2},z,r)\vert^{2}\sigma^{2}(x,r)
\,\,.
\label{eq:2.1}
\endarr
Here $\vec{r}$ is the transverse separation of the quark and
antiquark in the photon, $z$ and $(1-z)$ are partitions of
photon's lightcone momentum between the quark and antiquark,
$\sigma(x,r)$ is the dipole cross section for interaction of
the $q\bar{q}$ dipole with
the proton target (hereafter we use $\sigma(x,r)$ of
Refs.~[11,12]), and the dipole distribution in the
transverse polarized photon
$\vert\Psi_{\gamma^{*}}(Q^{2},z,r)\vert^{2}$ derived in [7],
equals
\beq
\vert\Psi_{\gamma^{*}}(Q^{2},z,r)\vert^{2}
={6\alpha_{em} \over (2\pi)^{2}}
\sum_{i}^{N_{f}}e_{i}^{2}
\{[z^{2}+(1-z)^{2}]\varepsilon^{2}K_{1}(\varepsilon r)^{2}+
m_{q}^{2}K_{0}(\varepsilon r)^{2}\}\,\,,
\label{eq:2.2}
\endeq
where $\alpha_{em}$ is the fine structure constant, $e_{i}$ is the
quark charge in units of the electron charge, $m_{q}$ is the quark
mass,
\beq
\varepsilon^{2} = z(1-z)Q^{2}+m_{q}^{2}
\label{eq:2.3}
\endeq
and $K_{\nu}(x)$
is the modified Bessel function. The mass spectrum for the $q\bar{q}$
excitation was calculated in [4] and steeply decreases with $M^{2}$:
\beq
\left.{d\sigma_{D}\over dM^{2}dt}\right|_{t=0}
\sim    {M^{2}\over (Q^{2}+M^{2})^{3}  }\, .
\label{eq:2.4}
\endeq

The $\propto 1/M^{2}$ component of the mass spectrum comes from
the diffraction excitation of the $q\bar{q}g$ Fock state of the
photon containing the soft gluon which carries the fraction $z_{g}
\ll 1$ of photon's lightcone momentum and gives rise to a large
excited mass $M^{2} \propto Q^{2}/z_{g}$.
Let $\vec{r}, \vec{\rho}_{1},
\vec{\rho}_{2}$ be the $\bar{q}$-$q$, $g$-$q$ and $g$-$\bar{q}$
separations in the impact parameter (transverse size)
plane, $\vec{\rho}_{2}=
\vec{\rho}_{1}-\vec{r}$. Then, in the triple-pomeron regime of
$x_{\Pom},\beta \ll 1$,
\arr
 (Q^{2}+M^{2})
\left.{d\sigma_{D}
\over dt dM^{2}}\right|_{t=0}=
{}~~~~~~~~~~~~~~~~~~~~~~~~\nonumber \\
\int dz d^{2}\vec{r}d^{2}\vec{\rho}_{1}\,\,
\left\{
z_{g}|\Phi(\vec{r},\vec{\rho}_{1},\vec{\rho}_{2},z,z_{g})|^{2}
\right\}_{z_{g}=0} \cdot {
\sigma_{3}^{2}(x_{\Pom},r,\rho_{1},\rho_{2})-
\sigma^{2}(x_{\Pom},r) \over 16\pi}\,
\label{eq:2.5}
\endarr
in which the square of the 3-parton wave function $|\Phi|^{2}$
equals [5,6]
\arr
|\Phi(\vec{r},\vec{\rho}_{1},\vec{\rho}_{2},z,z_{g})|^{2}=
{}~~~~~~~~~~~~~~~~~~~~~~~~~~~~~~~~~~~~\nonumber\\
{1 \over z_{g}} {1 \over 3\pi^{3}}
|\Psi_{\gamma^{*}}(Q^{2},z,r)|^{2}\mu_{G}^{2}
\left|
g_{S}(R_{1})K_{1}(\mu_{G}\rho_{1}){\vec{\rho}_{1}\over \rho_{1}}
-g_{S}(R_{2})K_{1}(\mu_{G}\rho_{2}){\vec{\rho}_{2}\over
\rho_{2}}\right|^{2} \, .
\label{eq:2.6}
\endarr
Here $g_{S}(r)$ is the running color charge, $\alpha_{S}(r)=
g_{S}(r)^{2}/4\pi$, the arguments of color charges
are $R_{i}={\rm min}\{r,\rho_{i}\}$.
In the wave function (\ref{eq:2.6}), the $\mu_{G}K_{1}(\mu_{G}r)
\vec{\rho}/\rho$ emerges as $\vec{\nabla}_{\rho}K_{0}(\mu_{G}\rho)$,
where
$K_{0}(\mu_{G}\rho)$ is precisely the two-dimensional Coloumb-Yukawa
screened potential. This makes self-explanatory the
interpretation [5,6,8,9] of $R_{c}=1/\mu_{G}$
as the correlation (propagation) radius for perturbative gluons.
The 3-body interaction cross section equals [5,6]
\beq
\sigma_{3}(r,\rho_{1},\rho_{2})={9 \over 8}
[\sigma(\rho_{1})+\sigma(\rho_{2})] -
{1 \over 8}\sigma(r)                         \, \, .
\label{eq:2.7}
\endeq
Hereafter we focus on $x_{\Pom}=x_{\Pom}^{0}$ and, for sake
of brevity, $\sigma(r)$ stands for $\sigma(x_{\Pom}^{0},r)$. Notice
that the subtraction of $\sigma^{2}(r)$ in the integrand of
(\ref{eq:2.5}) corresponds to the renormalization of the
wave function of the $q\bar{q}$ state for the radiation of
perturbative gluons [5,6]. For a detailed discussion of the
consistency of the above formalism with
color gauge invariance constraints see [5,6].


\section{$A_{3\Pom}(Q^{2})$ in DIS: photon as an effective
gluon-gluon dipole}

Because of $K_{\nu}(z)\sim \exp(-z)$ at large $z$, and by virtue
of (\ref{eq:2.3}), the dipole distribution (\ref{eq:2.2}) gives
the typical size of the $q\bar{q}$ dipole
\beq
r^{2} \lsim R_{q\bar{q}}^{2}=
{1\over \varepsilon^{2} } \propto {1 \over Q^{2}}\, .
\label{eq:3.1}
\endeq
Consequently, at $Q^{2} \gg 1/R_{c}^{2}$, to the standard
leading Log$Q^{2}$ approximation,
the dominant contribution
to the diffraction cross section (\ref{eq:2.5}) comes from
\beq
r^{2} \ll \rho_{1}^{2}\approx\rho_{2}^{2} \sim R_{c}^{2} \, .
\label{eq:3.2}
\endeq
In the region (\ref{eq:3.2}), we have
\beq
\sigma_{3}(r,\rho_{1},\rho_{2})\approx {9\over 4}\sigma(\rho)
\gg \sigma(r) \, ,
\label{eq:3.4}
\endeq
where $\rho={1\over 2}(\rho_{1}+\rho_{2})$,
and the virtual photon interacts as an effective gluon-gluon dipole
of size $\rho \sim R_{c}$, with the $q\bar{q}$ pair acting as an
octet color charge. Also, in this region we have
\beq
\mu_{G}^{2}|K_{1}(\mu_{G}\rho_{1}){\vec{\rho}_{1}\over\rho_{1}}
-K_{1}(\mu_{G}\rho_{2}){\vec{\rho_{2}}\over \rho_{2}}|^{2} \simeq
{r^{2}\over \rho^{4}}{\cal F}(\mu_{G}\rho)\,,
\label{eq:3.5}
\endeq
so that the 3-parton wave function factorizes.
The form factor
${\cal F}(z) =z^{2}[K_{1}^{2}(z)+z K_{1}(z)K_{0}(z)
+{1\over 2} z^2 K_{0}^{2}(z)]$ satisfies ${\cal F}(0)=1$ and
${\cal F}(z) \propto\exp(-2z)$ at $z > 1$.

The resulting diffraction cross section (\ref{eq:2.5}) takes on
the factorized form
\arr
(Q^{2}+M^{2})\left.{d\sigma_{D} \over dt dM^{2}}\right|_{t=0}
=\int dz \,d^{2}\vec{r}\,\,
|\Psi_{\gamma^{*}}(Q^{2},z,r)|^{2}\cdot {16\pi^{2} \over 27}\cdot
\alpha_{S}(r) r^{2} \nonumber\\
\times {1\over 2\pi^{4}}\cdot\left({9\over 8}\right)^{3}
\cdot \int d\rho^{2}
\left[{\sigma(\rho)\over \rho^{2}}\right]^{2}
{\cal F}(\mu_{G}\rho) \,,
\label{eq:3.6}
\endarr
Making use of the fact that,
at moderately small $x_{\Pom}$ and $r^{2} \lsim R_{c}^{2}$,
for the proton target and 3 active flavors, the exchange
by two perturbative gluons gives [5,6]
\beq
\sigma(x_{\Pom},r) \approx {16\pi^{2} \over 27}r^{2}\alpha_{S}(r)
\log\left[{1\over \alpha_{S}(r)}\right]
\approx {16\pi^{2} \over 27}r^{2}\alpha_{S}(r) \, ,
\label{eq:3.3}
\endeq
modulo to the logarithmic factor $\sim \log[1/\alpha_{S}(Q^{2})]
\sim 1$, we have
\arr
\int dz \,d^{2}\vec{r}\,\,
|\Psi_{\gamma^{*}}(Q^{2},z,r)|^{2}\cdot {16\pi^{2} \over 27}\cdot
\alpha_{S}(r) r^{2} \approx \nonumber\\
\sigma_{tot}(\gamma^{*}p)=
\int dz \,d^{2}\vec{r}\,\,
|\Psi_{\gamma^{*}}(Q^{2},z,r)|^{2}\sigma(r)\, .
\label{eq:3.7}
\endarr
Consequently,
(\ref{eq:3.6}) becomes equivalent to (\ref{eq:1.5}) [we will be back
to a detailed calculation of $A_{3\Pom}(Q^{2})$ in section 5]
with
\beq
A_{3\Pom}(Q^{2}) \sim    A_{3\Pom}^{*}=
{1\over 2\pi^{4}}\cdot\left({9\over 8}\right)^{3}
\cdot \int d\rho^{2}
\left[{\sigma(\rho)\over \rho^{2}}\right]^{2} F(\mu_{G}\rho)\, .
\label{eq:3.8}
\endeq
As the factorization (\ref{eq:3.5}) holds simultaneously
for all the $q_{i}\bar{q}_{i}g$ states, we predict
independence of $A_{3\Pom}(Q^2)$ on the flavour $"i"$.
$A_{3\Pom}^{*}$ is dominated by $\rho \sim R_{c}$. Making use
of (\ref{eq:3.3}), we obtain the order
of magnitude estimate of $A_{3\Pom}^{*}$:
\beq
A_{3\Pom}^{*} \sim {1\over 16}R_{c}^{2}\sim 0.1~{\rm GeV}^{-2}\,.
\label{eq:3.9}
\endeq
Here the scale for $A_{3\Pom}^{*}$ is set by the size $\rho \sim
R_{c}$ of the $q\bar{q}g$ Fock state of the photon.
Following [8,9], here we have taken $\mu_{G}=0.75$\,GeV as suggested
by lattice QCD studies [10].

Eq.~(\ref{eq:3.3}) describes the contribution to the dipole cross
section $\sigma(x_{\Pom},r)$ from the exchange by perturbative
gluons. This component $\sigma^{(pt)}(x,r)$ is a solution of the
generalized BFKL equation [5,6,8,9] and rapidly rises towards large
${1\over x}$ , dominating the observed growth and giving a good
quantitative description of the proton structure function at
HERA [11]. The phenomenological description of $\sigma(x,r)$
at large dipole sizes, $r\gsim R_{c}$, requires introduction of
the nonperturbative component $\sigma^{(npt)}(r)$ of the
dipole cross section [11-14], which is expected to have a weak
energy dependence (the scenario [11,12] introduces
an energy-independent
$\sigma^{(npt)}(r)$) and must be inferred from experimental
data.
Here we wish to recall that real photoproduction of the $J/\Psi$
and exclusive leptoproduction of the $\rho^{0}$ at $Q^{2}\sim
10$\,GeV$^{-2}$,
probe the
(predominantly nonperturbative) dipole cross section at $r\sim
0.5$\,fm$\lsim 2R_{c}$ [12-14]. Real, and weakly virtual $Q^{2}
\sim 10$\,GeV$^{2}$, photoproduction of the open
charm probes the (predominantly perturbative) dipole
cross section at $r\sim {1\over m_{c}}\sim {1\over 2}R_{c}$ [11,12].
The proton structure function $F_{2}^{p}(x,Q^{2})$ probes the
dipole cross section in a broad range of radii from $r\sim 1$\,fm
down to $r\sim 0.02$\,fm. Successful quantitative description
of the corresponding experimental data in [11-14] implies that we
have a reasonably good, to a conservative accuracy
$\lsim $(15-20)\%, understanding of the
dipole cross section at $r\sim R_{c}$ of the interest for
evaluation of $A_{3\Pom}^{*}$.
Quantitatively, at $r \sim R_{c}$
the dipole cross section $\sigma(x_{\Pom}^{0},r)$ receives
approximately $1:2$ contributions from the exchange by
perturbative gluons (\ref{eq:3.3}), and
from the nonperturbative component $\sigma^{(npt)}(r)$. The
numerical calculation with the dipole cross section of Ref.~[12]
gives
\beq
A_{3\Pom}^{*}=0.56\,{\rm GeV}^{-2}\,  ,
\label{eq:3.10}
\endeq
which is of the same order in magnitude as $A_{3\Pom}(Q^{2}=0)
\approx 0.16$\,GeV$^{2}$ from the real photoproduction analysis
[3].

\section{Real photoproduction: diffraction excitation of
"constituent" quarks}




At a first sight, the mechanism of diffraction dissociation of real
photons, $Q^{2}=0$, is quite different from the above in DIS.
For diffraction dissociation of real
photons, in the dipole distribution (\ref{eq:2.2})  the typical
size is large, of the hadronic scale,
\beq
r^{2} \sim R_{q\bar{q}}^{2} \approx {1\over m_{q}^{2}} \gg R_{c}^{2}
\label{eq:4.1}
\endeq
Here, following [12,15], for light flavours we use $m_{q}=
0.15$\,GeV. Such a choice of $m_{q}$ in the wave function
(\ref{eq:2.2}) gives, with the same dipole cross section
$\sigma(x,r)$, a good quantitative description of the real
photoabsorption cross section [12,4], of exclusive leptoproduction
of vector mesons at moderate $Q^{2}$ [14], of nuclear shadowing
in DIS on nuclei [15] and of color transparency effects in
exclusive production of vector mesons on nuclei [13,14].
Because of (\ref{eq:4.1}), the 3-parton distribution
$|\Phi|^{2}$ will be dominated by configurations with $\rho_{1}^{2}
\lsim R_{c}^{2} \ll \rho_{2}^{2}\sim r^{2}$ and $\rho_{2}^{2}
\lsim R_{c}^{2} \ll \rho_{1}^{2}\sim r^{2}$.
The dipole distribution
in the $q\bar{q}g$ state takes on
the factorized form first considered in [4]:
\arr
|\Phi (\vec{r},\vec{\rho}_{1},\vec{\rho}_{2},z,z_{g})|^{2}=
{}~~~~~~~~~~~~~~~~~~~\nonumber\\
{1 \over z_{g}} {4 \over 3\pi^{2}}
|\Psi_{\gamma^{*}}(Q^{2},z,r)|^{2}\mu_{G}^{2}
\left[ \alpha_{S}(\rho_{1})
K_{1}^{2}(\mu_{G}\rho_{1})+
\alpha_{S}(\rho_{2})K_{1}^{2}(\mu_{G}\rho_{2})\right]\, .
\label{eq:4.2}
\endarr
We recover a sort of
the constituent quark model, in which the gluon clusters with the
(anti)quark into the $qg$ and/or $\bar{q}g$ cluster of size
$\rho \sim R_{c}$, with the square of the $qg$
wave function of the
"constituent" quark $\propto
{1\over z_{g}}\alpha_{S}(\rho)K_{1}^{2}(\mu_{G}\rho)$\, .
 Diffraction dissociation of the photon
$\gamma^{*} \rightarrow q+\bar{q}+g$ proceeds via diffraction
excitation of the "constituent"
(anti)quark $q(\bar{q}) \rightarrow q(\bar{q})+g$
of the parent $q\bar{q}$ state of the photon.
Eq.~(\ref{eq:2.7})
leads to the estimate $\sigma_{3}(r,\rho_{1},\rho_{2})
\approx \sigma(r)+{9\over 8}\sigma(\rho)$ and
\beq
\sigma_{3}^{2}(r,\rho_{1},\rho_{2}) - \sigma^{2}(r)\approx
{9\over 4}\sigma(r)\sigma(\rho) \, ,
\label{eq:4.3}
\endeq
where $\rho={\rm min}\{\rho_{i}\}$. Then, the diffraction cross
section (\ref{eq:2.5}) takes the factorized form
\arr
M^{2}\left.{d\sigma_{D} \over dt dM^{2}}\right|_{t=0}
\approx\int dz \,d^{2}\vec{r}\,\,
|\Psi_{\gamma^{*}}(Q^{2}=0,z,r)|^{2}\sigma(r)
\nonumber\\
\times {3\over 8\pi^{2}}
\cdot \int d\rho^{2}
\left[{\sigma(\rho)\over \rho^{2}}\right] f^{2}(\mu_{G}\rho)=
\sigma_{tot}(\gamma p)A_{3\Pom}(0)
\,,
\label{eq:4.4}
\endarr
where
\beq
\sigma_{tot}(\gamma p)=
\int dz \,d^{2}\vec{r}\,\,
|\Psi_{\gamma^{*}}(Q^{2},z,r)|^{2}\sigma(r)
\label{eq:4.5}
\endeq
and
\beq
A_{3\Pom}(0) \approx
{3\over 8\pi^{2}}
\cdot \int d\rho^{2} \alpha_{S}(\rho)
\left[{\sigma(\rho)\over \rho^{2}}\right]  f^{2}(\mu_{G}\rho)
\sim {1\over 18}R_{c}^{2}\, ,
\label{eq:4.6}
\endeq
where $f(z)=zK_{1}(z)$.
A comparison of estimates (\ref{eq:3.9}) and (\ref{eq:4.6}) shows
that, modulo to the numerical, and logarithmic, factors $\sim 1$,~
we obtained
$A_{3\Pom}(Q^{2}) \approx A_{3\Pom}(0)$ , which was conjectured
some time ago in [4-7,16]. As a matter of fact, the exact large-$r$
behaviour of $\sigma(r)$ is not a main point here, we only should
assume that (\ref{eq:4.5}) reproduces the observed total
phototabsorption cross section. With the dipole cross section of
Ref.~[12] we find $\sigma_{tot}(\gamma p)=108$\,$\mu$b, in good
agreement with the Fermilab data [17]. A direct calculation from
(\ref{eq:2.5})-(\ref{eq:2.7}) gives
\beq
A_{3\Pom}(0)=0.23{\rm GeV^{-2}}
\label{eq:4.7}
\endeq
in agreement with the real photoproduction determination
$A_{3\Pom}\approx 0.16$\,GeV$^{-2}$ [3].


\section{$Q^{2}$ dependence of $A_{3\Pom}(Q^{2})$. Discussion of
results.}




To have more insight into the $Q^{2}$ dependence of the
triple-pomeron coupling, here we present the results of a direct
evaluation of $A_{3\Pom}(Q^{2})$ from equations
(\ref{eq:1.5}) and (\ref{eq:2.5}). We consider $x_{\Pom}=0.03$ and
in the case of DIS, we take $x=0.004$. In this range of $x$
and moderate $Q^{2}\lsim 10$\,GeV$^{2}$, the proton structure
function is approximately flat {\sl vs.} ${1\over x}$.
The equivalent {\sl c.m.s.} energy in the real photoproduction
can be estimated
as $W^{2} \sim m_{p}^{2}/x \sim 250$\,GeV$^{2}$, which
 corresponds to
the energy range of the Fermilab experiment [3].
We calculate the corresponding real and virtual photoabsorption
cross section from Eqs.~(\ref{eq:3.7},\ref{eq:4.5}) (for a
detailed comparison with experiment  see [11,12,14]).

Our results for the $Q^{2}$ dependence of the triple-pomeron
coupling $A_{3\Pom}(Q^{2})$ are presented in Fig.1. The main
feature of $A_{3\Pom}(Q^{2})$ is its weak $Q^{2}$ dependence,
which was anticipated in [4-7,16]. Still, we predict
a slight, by a factor $\sim 1.6$, growth of $A_{3\Pom}(Q^{2})$
from $A_{3\Pom}(Q^{2}=0)=0.23$\,GeV$^{2}$
to the DIS value
$A_{3\Pom}(Q_{\Pom}^{2} \approx 10 GeV^2)=0.36$\,GeV$^{-2}$.
This rise of $A_{3\Pom}(Q^{2})$ is predicted to take place
predominantly at $Q^{2} \lsim Q^{*2}\approx$(2-3)\,GeV$^{2}$.
The scale for $Q^{*2}$ corresponds to the transition from the
regime of diffraction dissociation of the "constituent" quark
of section 4 to the regime of diffraction dissociation of the
octet-octet state of the photon of section 3. This transition takes
place when the typical size of the $q\bar{q}$ pair $R_{q\bar{q}}$
becomes of the order of the size $\sim R_{c}$ of the "constituent"
quark:
\beq
R_{q\bar{q}} = {1\over \varepsilon} \approx {2\over \sqrt{Q^{2}}}
\sim R_{c},
\label{eq:5.1}
\endeq
which
leads to the estimate
\beq
Q^{*2} \sim {4\over R_{c}^{2}} = 3\,{\rm GeV}^{2} \, .
\label{eq:5.2}
\endeq
The predicted $Q^{2}$ dependence of $A_{3\Pom}(Q^{2})$ can be
tested at HERA.
The onset of the GLDAP evolution of the pomeron structure
function requires $R_{q\bar{q}}^{2} \ll R_{c}^{2}$, {\sl i.e.,}
$Q^{2}\gsim Q_{\Pom}^{2} \gg Q^{*2}$. Following the analysis
[11],
a reasonable choice for the corresponding factorization scale is
$Q_{\Pom}^{2}=10$\,GeV$^{2}$. The value of $A_{3\Pom}(Q_{\Pom}^{2})$
determines the normalization of the input sea structure function
of the pomeron (more detailed analysis of the partonic
structure of the pomeron and of its evolution properties is
presented elsewhere [18]). Notice that for diffraction excitation
of "constituent" quarks in real photoproduction, the calculated
$A_{3\Pom}(0)$ is a linear functional of $\sigma(R_{c})$, whereas
for diffraction excitation of the octet-octet state of the photon
in DIS, $A_{3\Pom}(Q_{\Pom}^{2})$ is a quadratic functional of
$\sigma(R_{c})$. Hence the conservative $\sim$(15-20)\% uncertainty
in our present knowledge of $\sigma(R_{c})$ implies the conservative
theoretical uncertainty of $\sim $(15-20)\% and
$\sim$(30-40)\% in the predicted value of $A_{3\Pom}(0)$ and
$A_{3\Pom}(Q_{\Pom}^{2})$, respectively.

The indirect experimental evidence
for weak $Q^{2}$ dependence of $A_{3\Pom}(Q^{2})$ comes from
nuclear shadowing in DIS on nuclei. Diffractive excitation of
large masses contributes significantly to nuclear shadowing
at $x\ll 10^{-2}$, and in [15] it was shown
that calculations using the above photoproduction value of
$A_{3\Pom}(0)$ are in good agreement with the experiment.
Crude evaluations [4,19] of the total rate of diffractive DIS,
using the photoproduction value of $A_{3\Pom}(0)$, are also
consistent with the HERA data [20] (a detailed treatment of
diffractive DIS in the dipole approach to the BFKL
pomeron is presented elsewhere [18]).

The so-called absorption corrections, not considered here, will
slightly reduce $A_{3\Pom}(Q^{2})$ (for the relevant formalism
see [5,6]). The absorption correction is typically of the order
of $\sigma_{3}/8\pi B$, where $B$ is the diffraction slope.
In real photoproduction, $\sigma_{3}\sim \sigma(r\sim{1\over
m_{q}})$ and the absorption correction to $A_{3\Pom}(0)$ will
be larger than to $A_{3\Pom}(Q_{\Pom}^{2})$, where
$\sigma_{3} \sim {9\over 4}\sigma(r\sim R_{c})$ is much
smaller. Therefore, the increase of $A_{3\Pom}(Q^{2})$
by the factor $\sim 1.6$ cf. $A_{3\Pom}(0)$
will be retained and, as a matter of fact somewhat enhanced, by
the absorption corrections.

To summarize, we presented the first direct evaluation of the
triple-pomeron coupling $A_{3\Pom}$. With $R_{c}\approx 0.3$\,fm ,
we find good agreement with the experimental determination of
$A_{3\Pom}(0)$. We related the small numerical value of
$A_{3\Pom}$, which was a mystery, to the
small correlation radius $R_{c}$ for the perturbative gluons.
We predict a slight rise of $A_{3\Pom}(Q^{2})$ with $Q^{2}$, by
a factor $\sim 1.6$, from real photoproduction to DIS at
$Q^{2}\sim 10$\,GeV$^{2}$.

In a somewhat related approach to the BFKL pomeron, the
triple-pomeron regime was considered also in [21]. These authors
consider the scaling BFKL regime of $R_{c}=\infty$ and fixed
$\alpha_{S}=$const [22], and their results are not applicable to
the forward diffraction dissociation ($t=0$) considered here.
\medskip\\
{\bf Acknowledgements:} B.G.Zakharov thanks J.Speth for the
hospitality at the Institut f\"ur Kernphysik, KFA, J\"ulich.
This work was partially supported by the INTAS grant 93-239.

\pagebreak

\pagebreak
{\bf \Large Figure captions}
\begin{itemize}
\item[Fig.1]
 - Our prediction for the $Q^{2}$ dependence of $A_{3\Pom}(Q^{2})$.

\end{itemize}
\end{document}